\documentclass[12pt,preprint]{aastex}

\slugcomment{to appear in ApJ}

\shorttitle{Search for Deuterated PAH features}
\shortauthors{Onaka et al.}

\begin{document}


\title{Search for the Infrared Emission Features from Deuterated Interstellar Polycyclic
Aromatic Hydrocarbons}


\author{Takashi Onaka\altaffilmark{1}, 
Tamami I. Mori\altaffilmark{1},
Itsuki Sakon\altaffilmark{1},
Ryou Ohsawa\altaffilmark{1},
Hidehiro Kaneda\altaffilmark{2},
Yoko Okada\altaffilmark{3},
and
Masahiro Tanaka\altaffilmark{4}
}

\altaffiltext{1}{Department of Astronomy, Graduate School of Science, %
The University of Tokyo, %
7-3-1, Hongo, Bunkyo-ku, Tokyo, 113-0033, Japan}
\email{onaka@astron.s.u-tokyo.ac.jp}%
\altaffiltext{2}{Graduate School of Science, Nagoya University, %
Chikusa-ku, Nagoya, 464-8602, Japan}%
\altaffiltext{3}{I. Physikalisches Institut, Universit\"at zu K\"oln, 50937 K\"oln, Germany}
\altaffiltext{4}{Center for Computational Sciences, 
University of Tsukuba, Ibaraki 305-8577, Japan}


\begin{abstract}
We report the results of a search for emission features from interstellar deuterated polycyclic
aromatic hydrocarbons (PAHs) in the 4\,$\mu$m region with the Infrared
Camera (IRC) onboard {\it AKARI}.  No significant excess emission is seen in 4.3$-$4.7\,$\mu$m
in the spectra toward the Orion Bar and M17 after the subtraction of line emission from
the ionized gas.  A small excess of emission remains at around 4.4 and 4.65\,$\mu$m, but the
ratio of their intensity to that of the band emission from PAHs at 3.3$-$3.5\,$\mu$m is estimated as 2$-$3\%.
This is an order of magnitude smaller than the values
previously reported and also those predicted by the model of deuterium depletion onto PAHs. 
Since the subtraction of the ionized gas emission introduces an uncertainty, 
the deuterated PAH features are also searched for
in the reflection nebula GN 18.14.0, which does not show emission lines from ionized gas.  
We obtain a similar result  that 
excess emission in the 4\,$\mu$m region, if present, is about  2\% of  the PAH band emission in the
3\,$\mu$m region.
The present study does not find evidence for the presence of the large amount of
deuterated PAHs that the depletion model predicts. The results are discussed in the context of deuterium depletion in the
interstellar medium.
\end{abstract}



\section{Introduction}
Deuterium (D) is one of the light elements created in the big bang, whose 
primordial abundance depends sensitively on the cosmological constants
\citep{boesgaard85}.  D is destroyed by nuclear reactions in
stellar interiors, a process termed astration, and its abundance decreases monotonically 
along with the chemical evolution of the Galaxy \citep{epstein76, mazzitelli80}.
The abundance of D in the present day is thus directly related
to the primordial nucleosynthesis and the subsequent Galactic chemical evolution.
The D astration factor, defined as the ratio of the primordial to the 
present-day D to hydrogen (D/H) ratio, should reflect the history of the star-formation and 
the infall of the pristine gas to the Galactic disk
\cite[e.g.,][]{romano06, tsujimoto11}.
However, the observed D/H ratio of the interstellar gas
shows no systematic trend with the metallicity, but shows a considerable scatter, which cannot
be accounted for solely by Galactic chemical evolution models \citep{tosi10}.
\citet{linsky06} show several pieces of evidence that D is depleted onto dust
grains in the interstellar medium (ISM), which was originally suggested by \citet{jura82}. 
They show that the D/H ratio of the interstellar gas is
well correlated with the depletion of iron and silicon and also with the excitation temperature
of molecular hydrogen.  Enrichment of D found in interplanetary dust
particles  \citep{messenger02} further supports the depletion model of D.

\citet{draine06} proposes that interstellar polycyclic aromatic hydrocarbons (PAHs)
can be the major reservoir of interstellar D.  PAHs are thought to be responsible for
a series of the emission bands in the near- to mid-infrared region \citep{leger84, allamandola85, 
allamandola89}, referred to as ``PAH features,''
which are ubiquitously observed in the ISM \citep[e.g.,][]{onaka96, mattila96, tsumura13}.
PAHs are small particles of less than 1\,nm in size and carry about 3.5\% of cosmic carbon atoms
\citep{tielens08}.
Based upon a thermodynamic equilibrium argument, 
\citet{draine06} shows that the D/H ratio in PAHs can be as high as $\sim 0.3$, 
which is sufficient to 
account for the observed range of D/H of the interstellar gas,  if the gas temperature is lower than 90K.

When H in PAHs is replaced by D, the emission bands originating from vibration modes of
C\sbond H bonds should be shifted to longer wavelengths by a factor of the difference in reduced mass between
the C\sbond H and C\sbond D oscillators.  The emission bands at 3.3 and 3.4$-$3.5\,$\mu$m,
which come from stretching vibrations of aromatic and aliphatic C\sbond H bonds, are thus expected to 
move to $\sim 4.4$ and $\sim 4.6$$-$4.7\,$\mu$m in deuterated PAHs \citep[PADs,][]{hudgins04}.
\citet{verstraete96} report detection of an emission feature at 4.65\,$\mu$m with the 4.4$\sigma$ level in
M17 based on observations with the Short Wavelength Spectrometer (SWS) onboard the {\it Infrared Space Observatory (ISO)}.
Subsequently \citet{peeters04} report detection of emission bands at 4.4 and 4.65\,$\mu$m at 1.9$\sigma$
and 4.4$\sigma$ levels in the Orion Bar region with SWS observations and attribute them to aromatic and aliphatic
stretching vibration modes of C\sbond D in PADs, respectively.  They estimate the ratio of the sum of the integrated band
intensities at 4.4 and 4.65\,$\mu$m to those at 3.3 and 3.4$-$3.5\,$\mu$m as $0.17\pm0.03$ in 
the Orion Bar and $0.36\pm0.08$ in M17.  While those ratios are in the range of D/H 
predicted by the depletion model of D onto PAHs \citep{draine06}, the spectral range 4.08$-$5.30\,$\mu$m of the
SWS (band 2a) has a rather high noise level and similar features are not detected in any other objects.
Therefore, it is strongly desirable to obtain spectra with higher signal-to-noise ratios and confirm the detection,
particularly for the 4.4\,$\mu$m band, giving a better constraint on the D/H ratio in PAHs.

The spectral range of 4.3$-$4.7\,$\mu$m is largely obscured by the
terrestrial atmosphere and we need space telescopes to observe these PAD features.
The Infrared Camera (IRC) onboard the {\it AKARI} infrared satellite \citep{murakami07}
has a spectroscopic capability in the near-infrared (NIR; 2$-$5\,$\mu$m) 
with a high sensitivity \citep{onaka07, ohyama07}.
In {\it this paper}, we report observations of the PAD features in the Orion Bar, M17, and the reflection
nebula GN18.14.0
in 4.3$-$4.7\,$\mu$m with the Infrared Camera (IRC), carried out in its warm mission phase
 \citep[Phase 3,][]{onaka10}.

\section{Observations and data reduction}
The present observations were carried out in the warm mission phase of {\it AKARI} after the
exhaustion of liquid helium, where only NIR observations were executed, as part
of the ``Interstellar Medium in our Galaxy and Nearby Galaxies'' program \citep[ISMGN,][]{kaneda09}.
Even after the exhaustion of liquid helium, the telescope and focal plane instruments were kept 
sufficiently cold ($<47$\,K) by the onboard cryocooler and the NIR channel of the IRC was able to be operated.
At temperatures below 47\,K, the thermal background from the telescope was still negligible
for wavelengths shorter than 5\,$\mu$m and the IRC was able to perform NIR spectroscopy with high
sensitivity compared to large, but warm telescopes, particularly for observations of the diffuse emission,
being free from the disturbance of the terrestrial atmosphere
\citep{onaka10}.  

The IRC has a variety of options for the NIR spectroscopy using its special aperture mask
pattern \citep{onaka07}.  In the present observations, 
slit spectroscopy was performed, using the Ns and Nh slits with the grism disperser.  The Ns slit is $0\farcm8 \times
5\arcsec$, while the Nh slit has a size of $1\arcmin \times 3\arcsec$.
The Ns and Nh slits observe the sky separated by about 2\arcmin\, simultaneously.
The Ns slit width is adjusted to the mid-infrared (MIR) channel and the Nh width matches with
about 2 pixels of the NIR channel (2\farcs92).
The NIR grism (NG) provides a dispersion of 0.0097\,$\mu$m per pixel for 2.5$-$5\,$\mu$m
and spectroscopy with Nh gives a slightly better spectral resolution than that with Ns.

\begin{figure}
\plotone{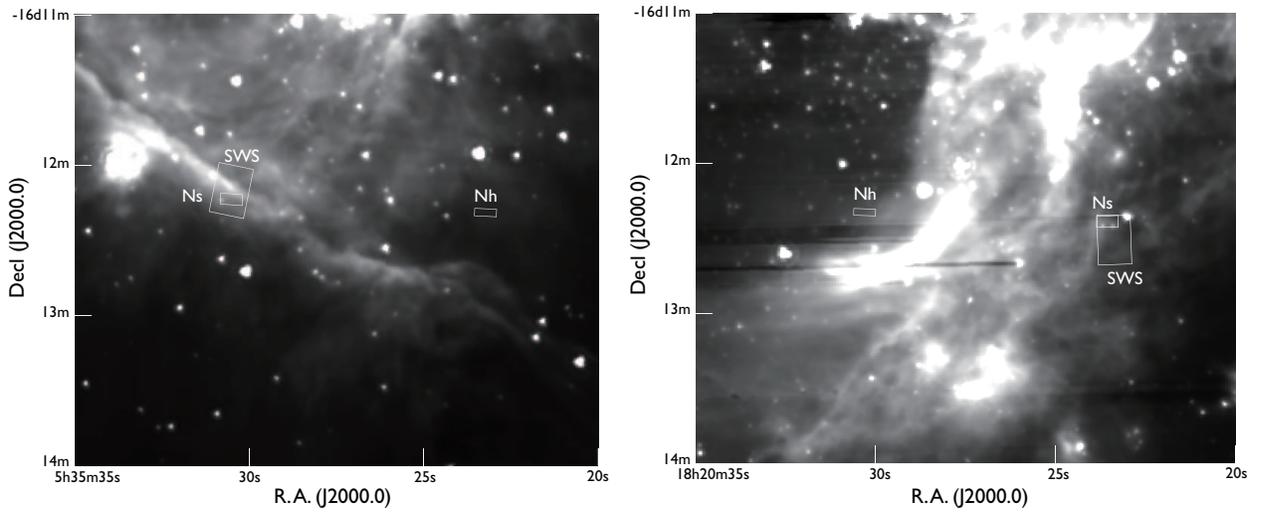}
\caption{Slit positions of (a) the Orion Bar and (b) M17.  The large
boxes indicate the SWS aperture positions and small boxes the
IRC Ns and Nh slit positions, where the present spectra are extracted.  
The background is the IRAC
band 2 (4.5\,$\mu$m) images retrieved from the NASA/IPAC Infrared Science Archive. \label{fig1}}
\end{figure}

In the program ISMGN, a number of extended objects in the Galactic plane have been observed
with NIR spectroscopy, including the Orion Bar and M17.
For both targets, the center of the Ns slit was intended to be at the position of the SWS
observations.  Since the IRC Ns slit is smaller than the aperture of the SWS ($20\arcsec \times 14\arcsec$), only
a part of the SWS aperture is covered with the Ns slit.
Due to the limited accuracy of the absolute pointing, the actual slit position was slightly
different from those intended, but the Ns slit overlapped with part of the SWS aperture in both targets.  
NIR Ns spectra are extracted from the overlapping region with the slit length of 6 pixels (8\farcs76) and
they are analyzed in the following sections.   Variation along the slit is not significant for both targets.
The central part of the Nh slit spectra is also extracted for comparison.
Figure~\ref{fig1} shows the slit positions of the SWS (large boxes) and the regions where the spectra are
extracted in the present study (small boxes).
The Nh slit turns out to be located toward the ionized region in both objects (see \S~\ref{sec3}).
In addition to these two objects, we also include the reflection nebula G18.14.0, which
does not show a signature of the ionized gas, in the present study.  
The observation was carried out toward the western boundary of the nebula of G18.14.0. 
Details of the present observation data are summarized in Table~\ref{table1}.

The data reduction was performed with the official pipeline optimized for Phase 3 version 20111121.
Due to the increase in the temperature in Phase 3, the dark current becomes non-negligible and
the number of hot pixels increases.  Even after the subtraction of the dark current in the official
pipeline, the residual dark signals are still recognizable at the edge region of a 10 pixel width 
next to the Nh slit, which is supposed to be blocked by the aperture mask.  The residual dark signals are estimated 
from this region and subtracted in the post pipeline process.   This process also corrects for 
spurious patterns parallel to the spatial direction due to the detector anomaly.  
The present targets are sufficiently bright and
this correction has only a small effect on their spectra.
Then the spectra are averaged for a $3 \times 3$ pixel region, both in the spectral and spatial
directions, with the official pipeline software.
This averaging process significantly reduces the noise level with a minimum cost to spectral
and spatial resolutions.

\section{Results and analysis\label{sec3}}
\begin{figure}
\includegraphics[angle=0,scale=.35]{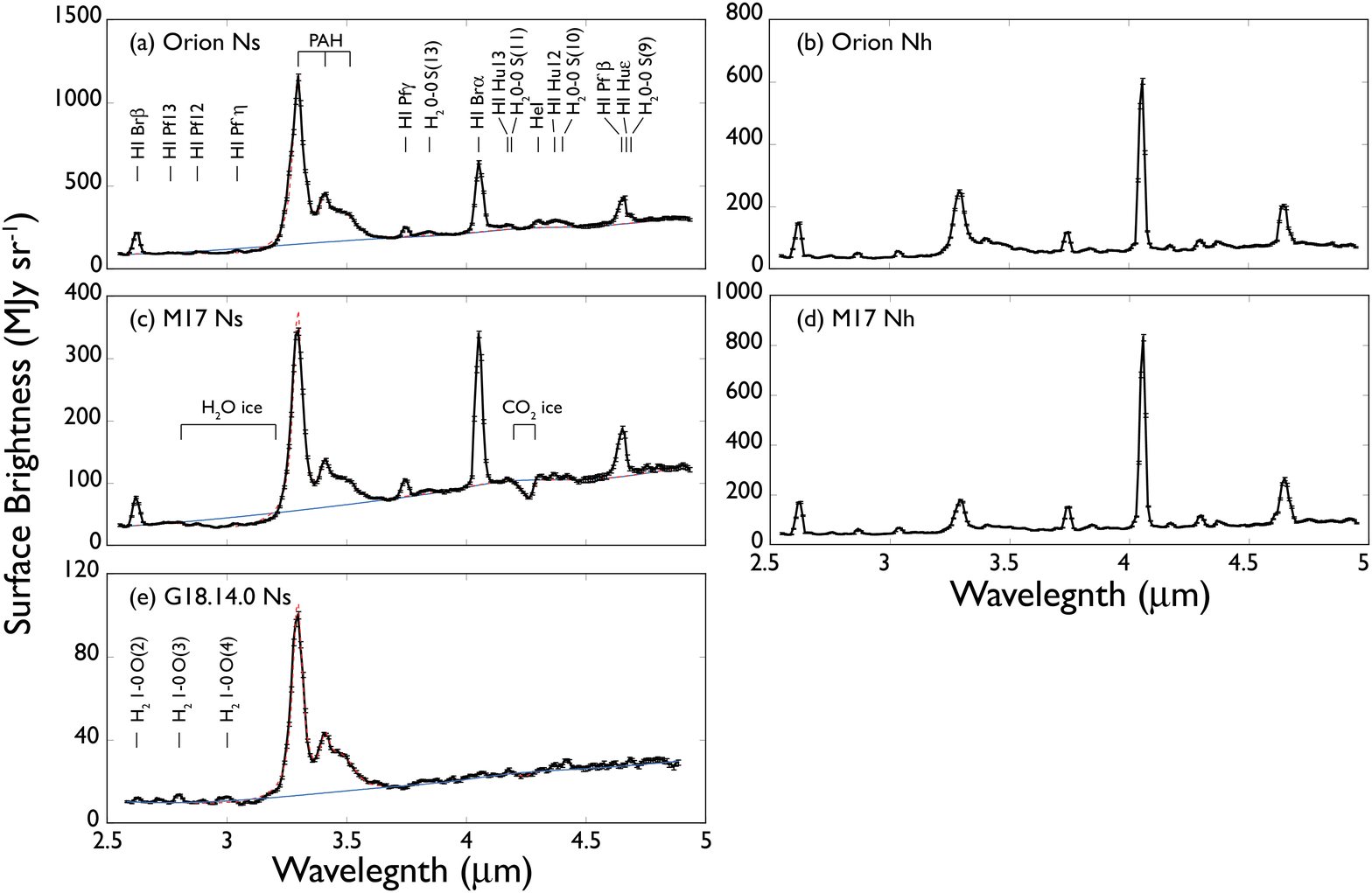}
\caption{IRC NIR spectra of (a) the Orion Bar at the Ns slit, (b) Orion region at the
Nh slit, (c) M17 at the Ns slit, (d) M17 at the Nh slit, and (e) G18.14.0 at the Ns slit.  In (a), (c), and (e) the assumed 
continua are indicated by the thin blue solid lines and the fitted spectra with the continuum, the 3\,$\mu$m
PAH band complex, and H$_2$O and CO$_2$ ice absorption are shown by the red dashed lines, which overlap with
most part of the observed spectra (see text).
Possible identification of the features is indicated in (a), (c), and (e).
Note that some faint features (e.g., Pf$\delta$ and H$_2$ S(12)) are not explicitly indicated 
since they are blended with strong emission features.
\label{fig2}}
\end{figure}

Figure~\ref{fig2} displays the obtained spectra.  The Ns spectra of the Orion Bar and M17 are rich in features.
Both are dominated by PAH band emission at 3.3$-$3.5\,$\mu$m and a number of hydrogen recombination
lines are also seen clearly.  The general characteristics of the Ns spectra are in agreement with
the SWS spectra \citep{peeters04} and are typical for the photodissociation region (PDR) associated with ionized gas.  
The Ns spectrum of M17 (Fig.~\ref{fig2}c) also indicates the presence of
CO$_2$ ice absorption at 4.27\,$\mu$m.  
The CO$_2$ ice absorption is not evident in other spectra.  
The hydrogen recombination lines appear quite strong in the Nh spectra of the Orion and M17 with
the faint PAH band emission, suggesting that the ionized gas dominates in these regions.  
The Ns spectrum of G18.14.0 shows clear PAH features, but the recombination lines
are not observed.  G18.14.0 is associated with a young stellar cluster \citep{bica03}.  
The spectral type of the central star of the nebula
HD\,167638 is B2II, consistent with the absence of ionized gas emission in the spectrum.
We assume that this target
represents a typical PDR without ionized gas.
Possible identification of the features in these spectra is indicated in Figures~\ref{fig2}a, c, and e
and summarized in Table~\ref{table2}.  Note that some of the features in these spectra are blended with bright
emission lines and bands, and therefore are not individually resolved.  Also unambiguous identification
of the faint features is difficult with the low spectral resolution of these observations.

The Ns spectra of the Orion Bar and M17 show a number of emission features in the range 4$-$5\,$\mu$m
(Figure~\ref{fig2}).  To estimate the feature intensities, first we make a spline fit to the continuum and
then fit the PAH band emission and the ice absorption.   
The pivot points for the spline fit are set as around 2.57, 2.67, 3.70, 3.95, 4.14, 4.50, and 4.88\,$\mu$m.
For the Orion Bar, the 4.14\,$\mu$m point is shifted to 4.24\,$\mu$m to avoid possible excess emission around
4.16\,$\mu$m.
Three PAH band components are clearly
seen at 3.29, 3.41, and 3.48\,$\mu$m.   The former two bands have sufficiently broad intrinsic widths compared to 
the present spectral resolution, while
the 3.48\,$\mu$m band is known to consist of more than one components, which cannot
be resolved by the present resolution \citep{geballe89}.  Hence, 
we approximate the 3.29 and 3.41\,$\mu$m bands
by Lorentzian functions and the 3.48\,$\mu$m band by a Gaussian in the fitting.
This combination of functions provides the best fit for the 3\,$\mu$m emission.
The CO$_2$ ice absorption at 4.27\,$\mu$m is clearly seen in the M17 spectrum.  
The presence of the CO$_2$ ice band also suggests the
presence of the broad H$_2$O ice absorption at around 3\,$\mu$m, 
 \citep[e.g.,][]{gibb04, shimonishi10}, but it is not obvious 
when the absorption is weak and overlaps with the bright PAH band emission.  
A simple screen geometry,
in which the ice species are located in front of the emission source,
is assumed in the spectrum fit.
For the H$_2$O ice absorption, we employ 
laboratory data taken at 10K \citep{ehrenfreund96}, whereas we assume a Gaussian profile in absorption
for the CO$_2$ ice because of its narrow width \citep{shimonishi10}. 
Details of the fit are given in \citet{mori13}, which employ a slightly different method to estimate the
continuum.  In the present study, we intend to fit the lower boundary of the spectrum in the
estimate of the continuum to obtain an upper limit for the remaining emission.

The best fit indicates the presence of H$_2$O ice with a column density of $(4.94 \pm 0.07) \times
10^{17}$\,cm$^{-2}$ for the Orion Bar, which is in the weakest end of the observed range \citep{gibb04}.
No positive detection of the CO$_2$ ice is obtained for the Orion Bar.  
The fit of the M17 spectrum provides the column densities of the H$_2$O and CO$_2$ ices
as $(7.65 \pm 0.07) \times 10^{17}$ and $(1.07 \pm 0.03) \times 10^{17}$ \,cm$^{-3}$, respectively.  
The ratio of the CO$_2$ to H$_2$O ice column densities is $14.0 \pm 0.4$\%, which is in agreement
with the ratios found in massive young stellar objects \citep[$17 \pm 3$\%,][]{gerakines99, nummelin01}.
The spectrum of G18.14.0 also suggests the presence of weak ice absorption.  The best fit yields a
column density of $(2.23\pm0.12) \times 10^{17}$ and $(0.32\pm0.06) \times 10^{17}$\,cm$^{-3}$ for
H$_2$O and CO$_2$ ices, respectively.  The column density ratio of $14\pm3$\% is again in the range expected
for massive young stellar objects.
The fitted spectra are shown by the red dashed lines in Figure~\ref{fig2}.

\begin{figure}
\epsscale{1.0}
\plotone{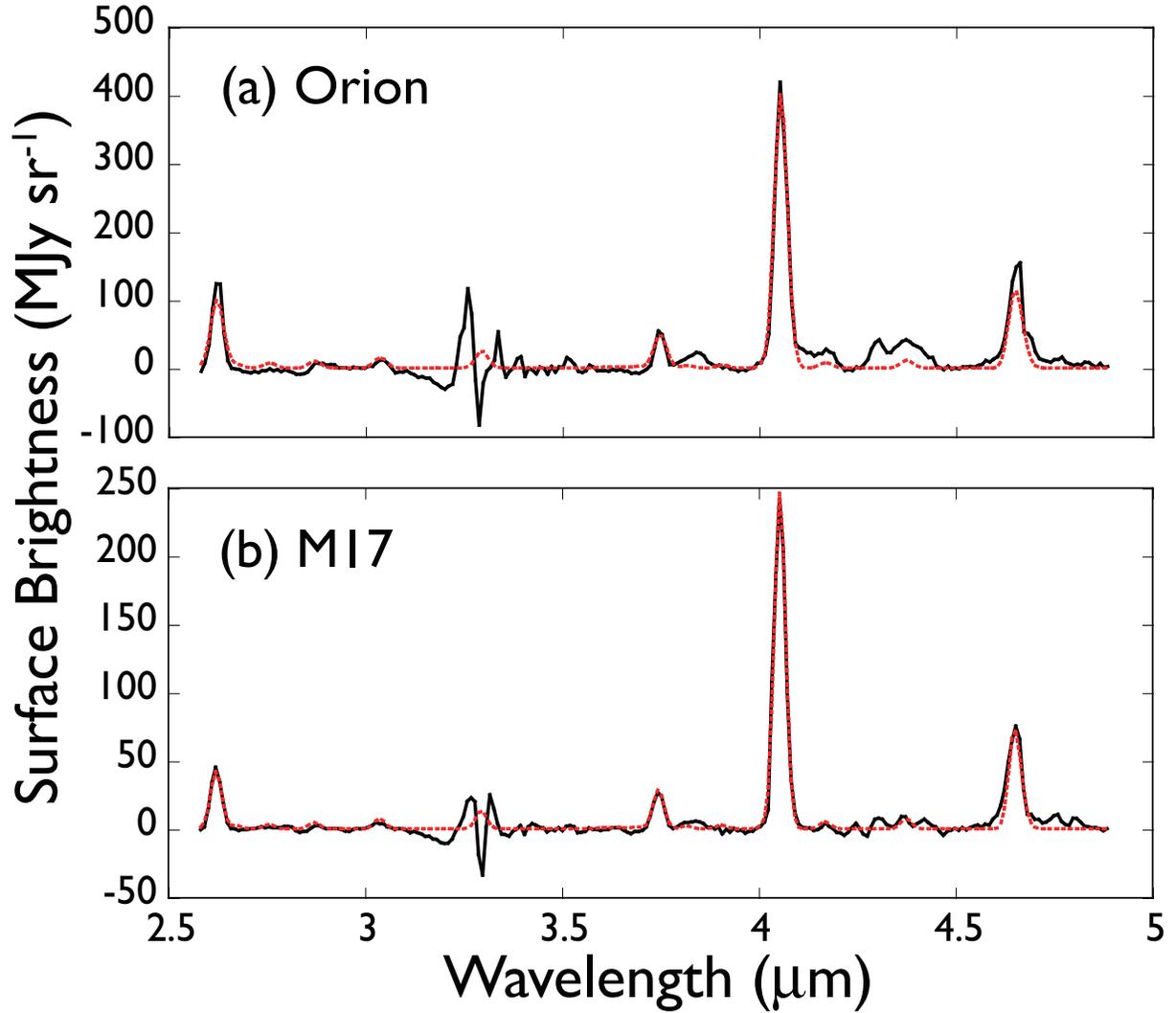}
\caption{Ns spectra from which the continuum, the PAH bands, and the ice absorption are subtracted for 
(a) the Orion Bar and (b) M17 (black solid lines).  The red dotted lines indicate the results of the model calculation
assuming the case B conditions with $T_e = 10^4$\,K and $n_e = 10^4$\,cm$^{-3}$ \citep{storey95}. \label{fig3}}
\end{figure}

In the next step we subtract the estimated
continuum, the PAH band emission, and the ice absorption from the observed spectrum.
Figures~\ref{fig3}a and b show the residual spectra for the Orion Bar and M17, which show the remaining emission 
features.  The residual seen around the 3\,$\mu$m region comes from the imperfect 
modeling of the PAH band emission, and it does not affect the following discussion.
In the range 4.3$-$4.7\,$\mu$m, several features appear clearly.  
The strongest one is \ion{H}{1} Pf$\beta$ at 4.6538\,$\mu$m.  Pf$\beta$ in the Orion spectrum
seems to be broader than other \ion{H}{1} lines and may be blended with the pure rotation line of 
H$_2$ S(9) at 4.6946\,$\mu$m.
 \ion{H}{1} Hu$\epsilon$ at 4.6725\,$\mu$m could also make a contribution.  
The 4.65\,$\mu$m emission of the Orion spectrum is found to be
better fitted by two Gaussians at 4.65 and 4.69\,$\mu$m, supporting the contribution from the
H$_2$ line (see below).
The presence of H$_2$ emission in the Orion spectrum
is also supported by possible detection of other pure rotational lines of H$_2$ of S(10), S(11), and
S(13) at 4.41, 4.18, and 3.85\,$\mu$m, respectively (Figure~\ref{fig2}a).  
The S(12) line overlaps with Br$\alpha$ and cannot be confirmed.
The H$_2$ lines are not clearly detected in the M17 spectrum.
The feature at 4.3\,$\mu$m, which appears strong in the Orion spectrum,
is identified as the recombination line of \ion{He}{1} $^3S_1$$-$$^3P_0$ \citep{vanden00}.
A broad emission is also seen at around 4.4\,$\mu$m.
The \ion{He}{1} line is weak and the presence of a feature around 4.4\,$\mu$m
is not obvious in the M17 spectrum.  There are no other excess emission features seen in
4.3$-$4.7\,$\mu$m.

\begin{figure}
\epsscale{0.9}
\plotone{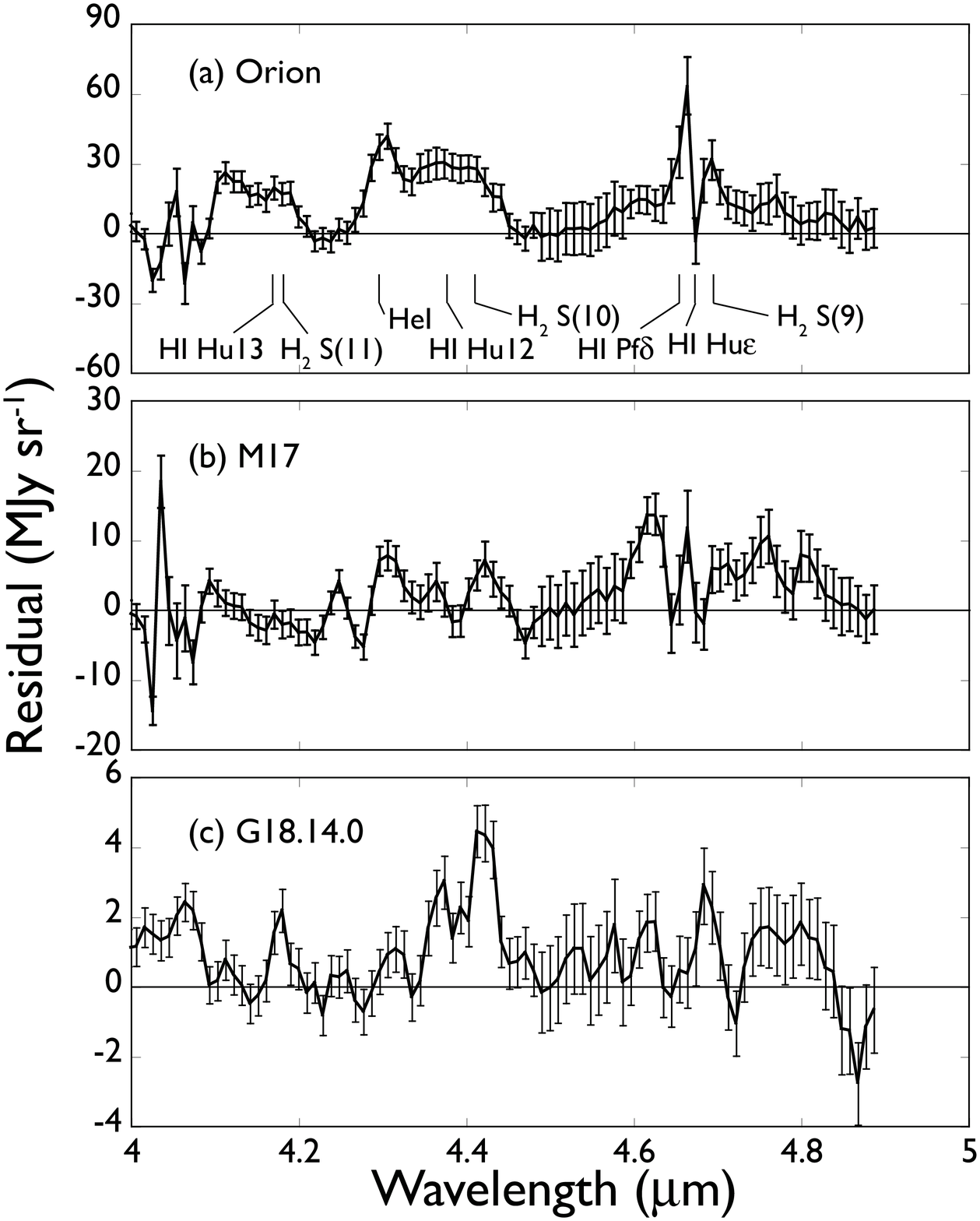}
\caption{Residuals of the subtraction of \ion{H}{1} recombination lines estimated with the Case B conditions for  
(a) the Orion Bar and (b) M17.  In (c), the residuals after subtraction of the continuum, the PAH band emission
and the ice absorption are plotted for G18.14.0.  In (a) the locations of the known lines are indicated.  
See text for details. \label{fig4}}
\end{figure}

To estimate the contribution of \ion{H}{1} recombination lines, we calculate a simple 
hydrogen recombination line model
of the Case B conditions with $T_e=10^4$\,K and $n_e = 10^4$\,cm$^{-3}$, where
$T_e$ and $n_e$ are the electron temperature and density of the ionized gas, respectively
\citep{storey95} and scale it to the observed Br$\alpha$ intensity.  
The estimated \ion{H}{1} emission is shown by the red dotted lines in
Figure~\ref{fig3}.   For the Orion spectrum we assume no extinction, while we set
$A_V = 10$\,mag for M17 to fit the Br$\beta$ intensity.   The models fit most of the \ion{H}{1} recombination
lines fairly well, including faint ones, particularly for M17, suggesting the validity
of the Case B calculation for the estimate of the \ion{H}{1} line intensities.
The uncertainty in $A_V$ is
estimated to be smaller than 1\,mag, which will affect the Pf$\beta$ intensity estimated from the 
Br$\alpha$ only by 0.7\%.
Then we subtract the model line emission from the observed spectra and
plot the residuals in the range 4$-$5\,$\mu$m in Figure~\ref{fig4}.  The
locations of the known emission lines (Table~\ref{table2}) are also indicated in Figure~\ref{fig4}a.
 The \ion{He}{1} line at 4.3\,$\mu$m remains
together with a broad emission around 4.4\,$\mu$m in the spectrum of the Orion Bar.  
Part of the 4.4\,$\mu$m residual emission can be attributed to H$_2$ S(10) and residual
\ion{H}{1}  Hu12, although the latter is too faint to account for the amount of the entire residual emission.
Residual emission is also seen
at 4.65$-$4.7\,$\mu$m.   Part of the residual comes from H$_2$ S(9) at 4.6946\,$\mu$m, 
which now appears clearly in Figure~\ref{fig4}a.
For M17, the residuals are smaller than the Orion spectrum.  There may be some residual in the blue
part of the Pf$\beta$ around 4.6\,$\mu$m.   

\begin{figure}
\epsscale{1.0}
\plotone{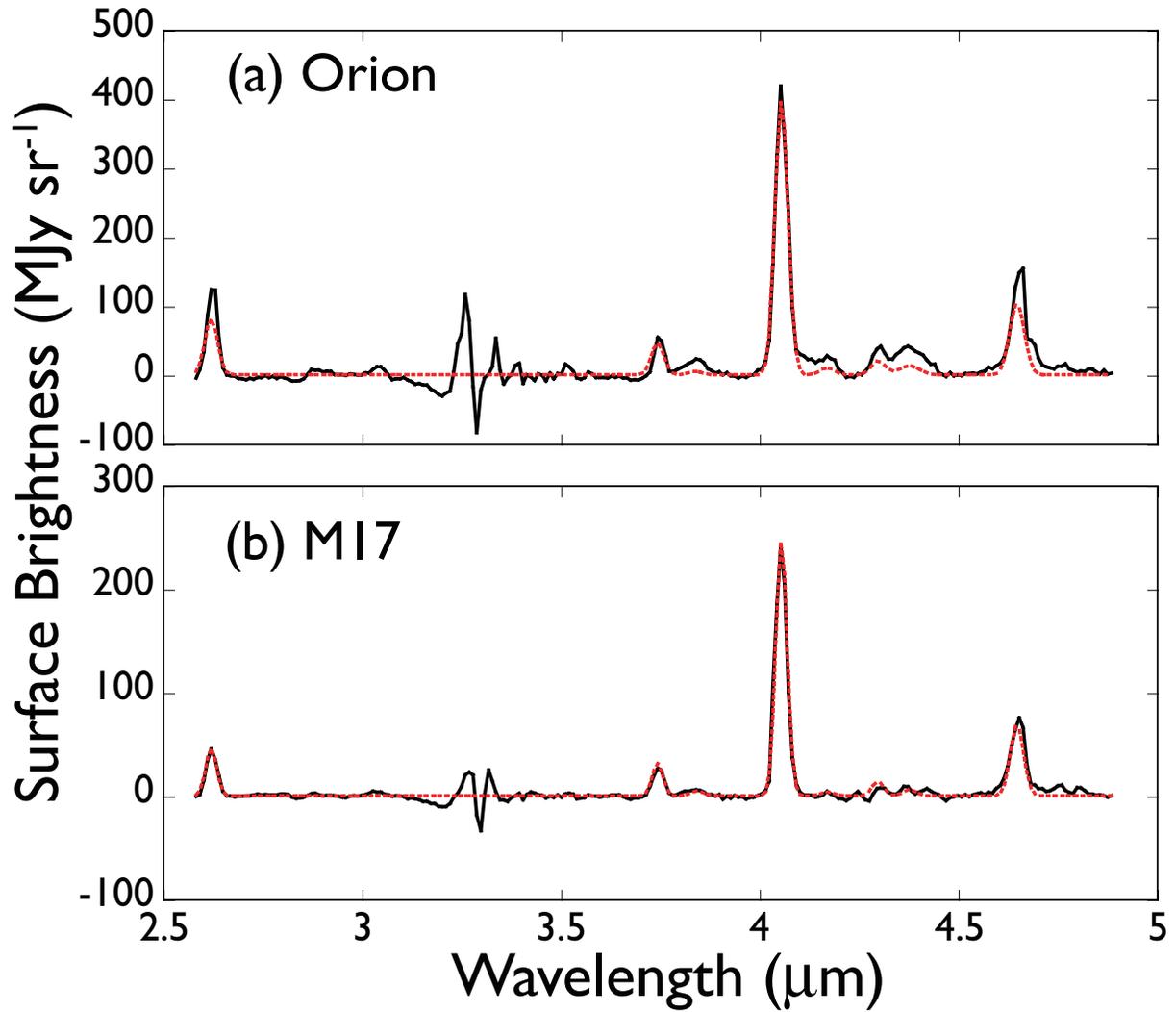}
\caption{The emission spectrum
estimated from the Nh spectrum and scaled to the Br$\alpha$ line (red dotted lines) for
(a) the Orion Bar and (b) M17.  
The black lines are the same as those in Figure~\ref{fig3} for
See text for details. \label{fig5}}
\end{figure}

The intensity of the He recombination line depends sensitively on the conditions of the ionized gas
in contrast to the hydrogen recombination lines and it is difficult to predict its intensity from the
present data accurately.  We cannot rule out a possibility
of unknown faint emission features present in the ionized gas either.  As an alternative way to
the Case B model calculation,
we make use of the spectra taken at the Nh slit, which are dominated by the
ionized gas, to estimate the emission lines from the ionized gas in each region. 
Here we simply assume that the relative emission line intensities do not vary from
the Nh to Ns positions appreciably as a first approximation.  Figure~\ref{fig5} shows the spectrum
estimated from the Nh spectra after being scaled to the Br$\alpha$ intensity and adjusted to the spectral
resolution of Ns.  The \ion{H}{1} emission line intensities
in the Nh spectra are also in good agreement with the Case B model calculation and thus the 
overall agreement
in the \ion{H}{1} lines is not unexpected.  

\begin{figure}
\plotone{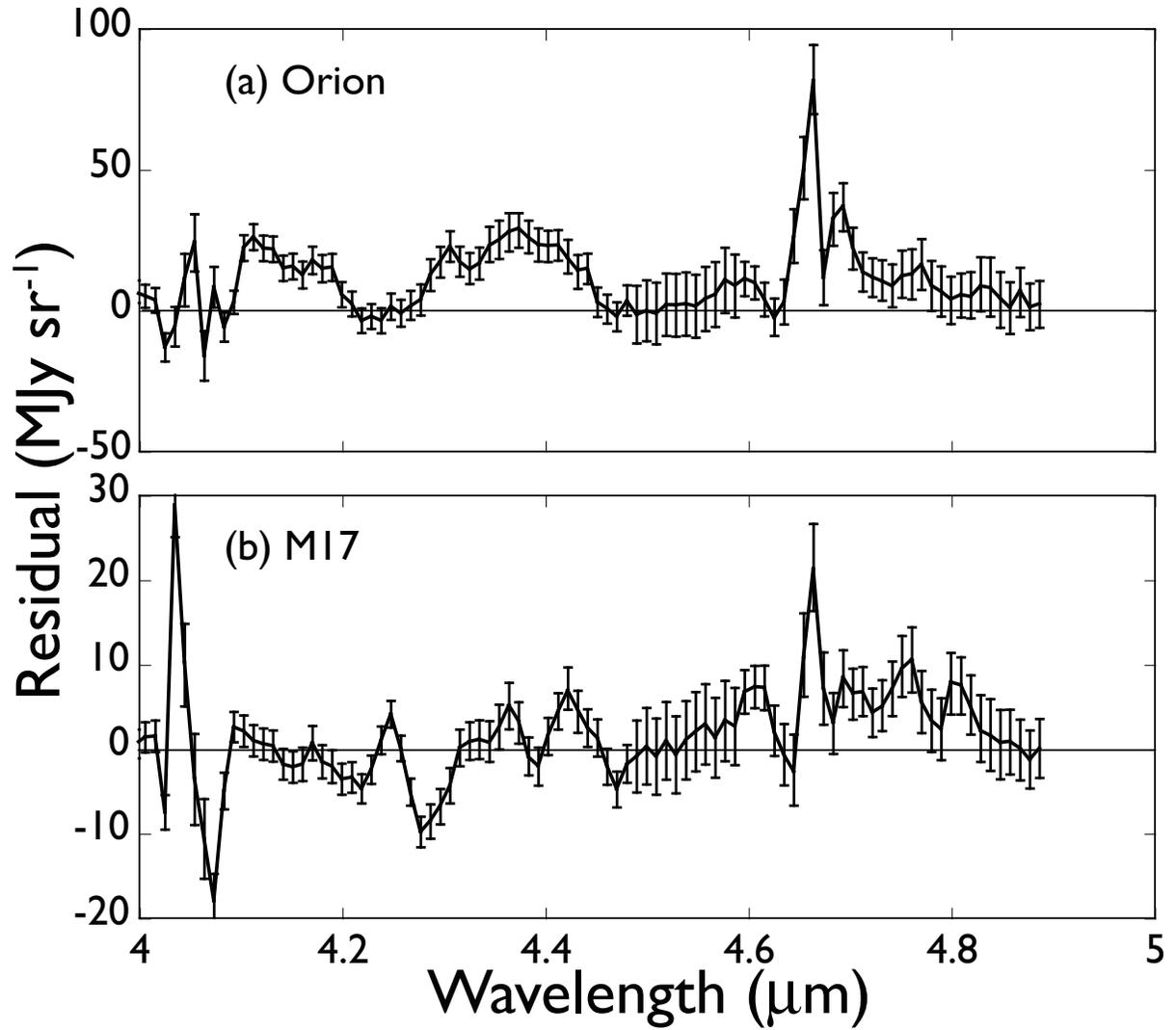}
\caption{Residuals of the subtraction of the emission spectrum estimated from the Nh spectrum.
(a) the Orion Bar and (b) M17.  See text for details. \label{fig6}}
\end{figure}

Figure~\ref{fig6} plots the residual between the observed Ns spectra and the spectra
estimated from the Nh spectra.  The residual emission at 4.3\,$\mu$m in the Orion spectrum is
reduced, confirming that it comes from the \ion{He}{1} line.  
The line emission estimated from the Nh spectra also accounts for part of the
emission at 4.4\,$\mu$m region in the Orion, but not perfectly, suggesting a partial contribution
from the ionized gas components.
The emission around 4.4\,$\mu$m
still remains.  The excess emission around 4.65\,$\mu$m is also reduced in the 
M17 spectrum to some extent.  Some excess emission is apparent at the position of Pf$\beta$
(4.65\,$\mu$m) in both spectra.

The residuals in Figure~\ref{fig6} are generally smaller than those in Figure~\ref{fig4}.
To estimate the excess intensities conservatively, however, we use the residuals shown in Figure~\ref{fig4}, 
in which the line emission estimated from the Case B conditions is subtracted.   We assume that
the emission at 4.3\,$\mu$m is the \ion{He}{1} line and take account of the emission only
at around 4.4\,$\mu$m.  
For the emission around 4.65\,$\mu$m, we separate the contribution
from H$_2$ S(9) by fitting two Gaussians for the Orion spectrum and estimate the
residual emission at around 4.65\,$\mu$m.  For the M17 spectrum, we simply calculate the
residual emission at 4.65\,$\mu$m.  
The results are summarized in Table~\ref{table3}
together with the intensities of the PAH bands.  
Intensities of the excess emission are estimated by a fit of the excess with a Gaussian and
errors are derived from the fitting, also taking account of any observational error.
For the estimate of the PAH bands, the separation
of the 3.41 and 3.48\,$\mu$m bands cannot be made accurately at this spectral resolution and only the summation of the
two features is given.   Contribution to the 3.29\,$\mu$m band  from Pf$\delta$ is subtracted
assuming the Case B model, which is small ($<3$\%)  for both spectra.  
In the following, we assume that the 3.29\,$\mu$m band comes from aromatic C\sbond H bonds
and the 3.41 and 3.48\,$\mu$m bands from aliphatic C\sbond H bonds in PAHs.  The features at
4.4 and 4.65\,$\mu$m are assumed to be the corresponding features from aromatic and aliphatic
C\sbond D bonds in PADs, respectively \citep{peeters04}.  No extinction correction is applied for the
estimate of the PAH band intensities.

The spectrum that contains contribution from the
ionized gas shows a number of emission lines.  It is difficult to predict and identify faint features
accurately with the resolution and sensitivity obtained here.  To avoid possible faint emission
from unknown species in the ionized gas, we require a spectrum that does not show a signature
of the ionized gas, but has PAH emission with a sufficient signal-to-noise in our targets.   Most of the spectra
taken in the ISMGN program show some sign of the ionized gas.
The reflection nebula G18.14.0 is the best target on these criteria in the ISMGN sample.  
The spectrum of G18.14.0 shows no evidence of the
ionized gas, but has prominent PAH emission.  There are weak emission features seen in 2.5$-$3\,$\mu$m
(Figure~\ref{fig2}e),
some of which can be attributed to H$_2$ 1$-$0 O(2), O(3), and (4) at 2.627, 2.803, and 3.004\,$\mu$m
\citep[e.g.,][]{lee11}.
The O(2) line is at the same wavelength as \ion{H}{1} Br$\beta$ and the absence of Br$\alpha$ suggests
that the excess at 2.63\,$\mu$m is attributable to O(2).  These bands are weak, but at least the O(3) line seems to be real.
Figure~\ref{fig4}c plots the residual after the subtraction of the continuum and
the PAH band emission with the ice absorption for 4$-$5\,$\mu$m (red dashed line in Figure~\ref{fig2}e).
There is  small excess emission remaining at around 4.4 and 4.65\,$\mu$m. 
The detection of H$_2$ 1$-$0 lines in 2.5$-$3\,$\mu$m suggests possible contribution from
H$_2$ 0$-$0 S(10) and S(9) to the excess at around 4.4 and 4.65\,$\mu$m, respectively.
The emission features in 2.5$-$3\,$\mu$m are too weak to constrain the emission mechanism and conditions with accuracy,
even if they are attributed to the H$_2$ emission.  Therefore, from the present spectra
it is rather difficult to estimate the intensities of H$_2$ S(9) and S(10) transitions accurately and thus
they are not subtracted from the intensity of the excess emission.
The intensities of the excess in G18.14.0 are also listed in Table~\ref{table3} together with the
PAH band intensities.  As indicated in Table~\ref{table3} the excess emission at 4.65\,$\mu$m
is only at the 2.5\,$\sigma$ level and the presence of any excess is marginal.

\section{Discussion}
\subsection{PAD to PAH ratio}
The last three columns of Table~\ref{table3} show the ratios of the intensities of PADs to those
of PAHs.  We estimate the ratios for the aromatic (4.4\,$\mu$m to 3.29\,$\mu$m) and
the aliphatic (4.65\,$\mu$m to 3.41 and 3.48\,$\mu$m) components
separately as well as for the total (aromatic + aliphatic)  PAD to PAH features.  The aliphatic component
in the Orion Bar and M17 shows the ratios of 4\% and 7\%, respectively, but the ratios of the aromatic component 
and the total intensities
are all less than 3\%.  The effect of extinction is not taken into account in these estimates.
If we take account for the extinction effect for M17, the ratio will be decreased further by about 20\%.
The extinction is estimated to be not significant for the Orion Bar.  For G18.14.0, the estimate of extinction 
using \ion{H}{1} lines
is not possible, but weaker ice absorption suggests that the effects should be smaller than for M17.
Note that the contributions from \ion{H}{1} lines are subtracted in these estimates, but
those from molecular hydrogen are not except for the S(9) line in the Orion Bar.

The present result shows a clear contrast to the results by \citet{peeters04}, who report the total ratio
of $0.17 \pm 0.03$ for the Orion Bar and $0.36\pm0.08$ for M17.  
The present spectra do not show any strong excess emission in the 4\,$\mu$m region except for \ion{H}{1} Pf$\beta$.
The remaining excess is about 10\% of Br$\alpha$ (Table~\ref{table3}) and the 
intensity of Pf$\beta$ is predicted with better accuracy than the excess.
This is also supported by the estimate using the Nh spectra.  
The basic result is
not affected by the method of the subtraction of the emission lines.  
The present observation covers only a part of the aperture of the SWS observations and 
it is possible that strong excess emission is present in the region that the IRC observations
did not cover.  The spatial variation along the Ns slit is, however, not significant, and no strong
excess emission in the 4\,$\mu$m region is seen in the entire slit spectra.  Thus it seems unlikely that 
the present observations miss the right spot in both targets, although we cannot rule out the possibility.  

It should also be noted that the present estimate gives upper limits on the PAD band emission.
The residual spectra, in which the
line emission estimated from the ionized gas (Nh) spectra is subtracted, show smaller residuals, particularly 
for the 4.4\,$\mu$m emission, than those after subtraction of the Case B spectra.  
There could be contributions from faint emission of unknown species
at around 4.4\,$\mu$m.  Marginal detection of the excess emission at 4.65\,$\mu$m for G18.14.0
also suggests that the residual emission in this spectral range in the Orion Bar and M17 may have
contribution from faint emission components in the ionized gas.   For G18.14.0, there may be contribution from 
the pure rotation lines of H$_2$.
Since these features are weak,
it is difficult to unambiguously attribute the excess emission seen
at 4.4 and 4.65\,$\mu$m, if real, to deuterated PAHs or others from the present data.
Spectra with higher spectral resolution and better
signal-to-noise ratios are needed to make clear confirmation and
identification of these excess emissions.

In the above discussion, we simply compare the intensities of the features 
in the 3 and 4\,$\mu$m regions  according to the
previous study \citep{peeters04}.  However, the emission intensity depends on
the transition probabilities, which may not be the same for C\sbond H and C\sbond D bonds.
Also the emission in the NIR requires high excitation, whose conditions are 
different between the emission at 3 and 4\,$\mu$m.   Bands at 4\,$\mu$m should be
more easily excited than those at 3\,$\mu$m.

\citet{bauschlicher97} show that 
perdueterated PAHs have integrated intensities for their spectroscopic transitions reduced by a factor of 1.75
compared to fully hydrogenated PAHs.  Assuming this reduction factor in the cross-section, we make a 
calculation based on a simple PAH emission model given by \citet{mori12} for
D/H = 0.1 and 0.025 to roughly estimate the effects.
The 3\,$\mu$m PAH emission is thought to come mainly from
neutral PAHs \citep{draine01}.  Since we discuss only the ratio of the 4.4\,$\mu$m to 3.3\,$\mu$m band
intensity, we simply assume that all the PAHs are neutral in the following estimate.
Even if a half of the PAHs are assumed to be ionized, the band ratio is changed only by 0.5\%.
We also assume as a standard case that the temperature
of the exciting source is 30000\,K, and the size distribution of PAHs is given by
a power-law with the number of carbon atoms in the 
smallest and largest PAHs of 20 and 4000, respectively.   
The results indicate
that the ratio of the 4.4\,$\mu$m to 3.3\,$\mu$m band intensity is about 0.095 and 0.024, respectively.
Easier excitation of
the 4.4\,$\mu$m band compensates the smaller cross-section, resulting in 
an intensity ratio similar to the D/H abundance ratio.   The result is obviously
sensitive to the assumed parameters of the exciting source and the size distribution of PAHs.
If the minimum number of carbon atoms is increased to 50, the intensity ratio becomes
0.14 and 0.035, respectively, which are larger than the abundance ratio by about 40\%.
Thus the observed intensity ratio may overestimate the true D/H ratio by several tens \%, but
not by an order of magnitude.  The present study does not find evidence for the D fraction
of more than 3\% in the carriers that emit the 3\,$\mu$m PAH band emission.

\subsection{Deuterium depletion model}
The expected amount of the depleted D onto solid particles depends on the abundance of
the observed distribution of the D/H ratios in the ISM, which is related to the primordial
D abundance via chemical evolution in the Galaxy.  The primordial
D/H can be estimated from observations
of distant quasars.   The latest estimate based on quasar observations suggests
(D/H)$_\mathrm{prim} = 25.35 \pm 0.05$\,ppm, which
is smaller than a previous estimate \citep{pettini08}, but is 
in good agreement with the most recent cosmological parameters deduced from the
angular power spectrum of the comics microwave background \citep{pettini12}.
\citet{prodanovic10} make a Bayesian analysis on the observations of gaseous D in the ISM and
find that the maximum and minimum D/H in the ISM are $20 \pm 1$ and $7 \pm 2$\,ppm, 
respectively, for a top hat distribution.  The astration factor suggested by these results becomes less than 1.3,
if the variation in the D/H ratio in the ISM is attributed to the depletion and the largest ratio is taken as the
intrinsic D/H at the present.  The small astration factor
can be accounted for by a recent model of the chemical evolution that takes account of the declining star-formation 
together with the infall of pristine gas onto the Galactic disk \citep{tsujimoto11}.

The above analysis of the interstellar D/H distribution suggests that D/H of about 13\,ppm 
needs to be depleted onto interstellar grains at maximum.  Using this value, 
the D/H ratio expected in dust grains is estimated following \citet{draine06}.
The C/H in interstellar grains is assumed to be about 200\,ppm, 85\% of which are in aromatic
\citep{pendleton02}.  For aromatic materials, H/C is about 0.35, which suggests
60\,ppm of H in aromatic grains.  Thus, the expected maximum D/H in PAHs is 0.28.  
The present result
does not find evidence for D/H $\ge 3$\% in PAHs and only accounts for the interstellar
D/H variation of $\sim 1$\,ppm.

UV observations of D in the ISM generally probe regions with medium density and cannot investigate
dense regions, while current NIR spectroscopy
is not sensitive enough to detect emission from regions with relatively small optical depth.
Thus if deuterated PAHs are present only in less dense regions, PAD emission cannot be detected
in the present observations.  Unimolecular photodissociation may work at moderate temperatures and can
increase D fraction in PAHs in regions where UV radiation penetrates
\citep{allamandola89, tielens97, hudgins04}.  However it is difficult to obtain the high concentration of D
suggested by the depletion observations
with this mechanism.  The observed amount of D depletion may be achieved only by the deuteration process
in dense and cold environments
\citep{draine06}.  If the depletion of D onto PAHs predominantly occurs in cold and dense regions
and is preserved in diffuse \ion{H}{1} clouds toward which UV observations are carried out, 
a signature of PADs should appear in PDRs, targets of the present observations.  

It should be noted that the present observation is only sensitive to the population of the smallest
aromatic dust grains
that emit 3$-$4\,$\mu$m emission.
If D is depleted in PAHs whose size is
too large to make prominent emission in the 3$-$4\,$\mu$m region, they could elude detection
in the NIR spectroscopy.   We estimate the size dependence using the simple model as above.
Assuming that only PAHs whose number of carbon atoms is larger than 50 are deuterated and
the total D/H = 0.1, then the 4.4 to 3.3\,$\mu$m band ratio becomes 0.049.  Hence, if a large amount of
interstellar D is depleted on PAHs, it might reside only in large PAHs.
Emission expected from large deuterated PAHs would appear at longer wavelengths
in a more complicated way than at the 4\,$\mu$m region
because of the interaction between C\sbond H and C\sbond D in bending
modes \citep{hudgins04}.  A dedicated search for MIR features from large PADs is needed to test the
depletion model and make a clear conclusion on D depletion onto PAHs.

\section{Summary}
We made sensitive spectroscopic observations in the NIR towards the Orion Bar, M17, and the reflection nebula 
G18.14.0 with
the IRC on board {\it AKARI}.  Neither of the spectra of the Orion Bar nor M17 shows evidence for strong emission features
expected from deuterated PAHs in the 4\,$\mu$m region.  Some excess emission remains
at around 4.4 and 4.65\,$\mu$m after subtraction of the contribution from hydrogen recombination lines,
but its intensity is less than previously reported.  
To avoid potential contributions from faint emission components in ionized gas,
the spectrum of the reflection nebula G18.14.0 is also analyzed in a similar way.
It shows a typical PDR spectrum without any sign of ionized gas.  
The spectrum indicates slight excess emission at around 4.4\,$\mu$m
and a marginal detection of the excess at 4.65\,$\mu$m (2.7$\sigma$).

These excess emissions are weak and it should be emphasized that
it is difficult to unambiguously rule out other possibilities and
attribute these two features to PADs because of the low
spectral resolution of the present observations.  
If we assume that they come from aromatic and aliphatic C\sbond D in PADs, then
the ratio of PAD (4.4 and 4.65\,$\mu$m) 
to PAH (3.29, 3.41, and 3.48\,$\mu$m) features is estimated as 3\% at most in the three targets.
If these 4\,$\mu$m features have different origins or there are  contributions from other
species, the ratio will further be reduced.

The effects of the difference in the cross-section and excitation between the 3 and 4\,$\mu$m emission
are investigated based on a simple model calculation.  The results suggest that the effects
may overestimate the D/H ratio by several tens \%, and do not affect the conclusion.  
The present study does not find evidence for the presence of a large amount of
deuterated PAHs as predicted from the D depletion model.

The PAD/PAH ratio required to account for the observed distribution of D in the ISM is larger by an order of 
magnitude
than the result of present observations.  If interstellar D is depleted onto PAHs, then it must
reside in large PAHs, which do not contribute dominantly to the NIR PAH emission.
Search for MIR emission from PADs is needed to further test the D depletion model onto 
PAHs, although the band emission of PADs at the MIR is more complicated than at the 4\,$\mu$m
region.

\acknowledgments

This work is based on observations with {\it AKARI}, a JAXA project
with the participation of ESA.  The authors thank all the members of
the {\it AKARI} project and the members of the Interstellar and Nearby
Galaxy team for their help and continuous encouragements.  They are also grateful to Bruce
Draine and Francois Boulanger for the
useful discussions and Mark Hammonds for careful reading of the manuscript.
This research has made use of the NASA/IPAC Infrared Science Archive, 
which is operated by the Jet Propulsion Laboratory, California Institute of Technology, 
under contract with the National Aeronautics and Space Administration.
This work is supported by Grants-in-Aid
for Scientific Research from the Japan Society for the Promotion of Science
(nos. 21654027 and 23244021).

{\it Facility:} \facility{AKARI}.

\clearpage

\begin{deluxetable}{ccrrcc}
\tablecaption{Journal of the observations \label{table1}}
\tablewidth{0pt}
\tablehead{
\colhead{Source}  & \colhead{Slit} & \colhead{$\alpha$ (J2000.0)} & \colhead{$\delta$ (J2000.0)} &  \colhead{Pointing ID} & \colhead{Observation date} }
\startdata
Orion Bar & Ns & 5h35m19.8s & -05\degr 25\arcmin 14\farcs2 & 1420483.1 & 2008 September 14 \\
                  & Nh & 5h35m13.0s & -05\degr 25\arcmin 19\farcs6 & 1420483.1 & 2008 September 14 \\\
M17          & Ns & 18h20m23.6s & -16\degr 12\arcmin 23\farcs2\ & 1420484.1   &  2008 September  27 \\             
                   & Nh & 18h20m30.3s & -16\degr 12\arcmin 19\farcs8 & 1420484.1   &  2008 September  27\\  
G18.14.0 & Ns & 18h16m58.6s  & -19\degr 47\arcmin 30\farcs0  & 1420707.1 & 2009 March 23\\      
\enddata
\end{deluxetable}

\clearpage

\begin{deluxetable}{ccl}

\tabletypesize{\scriptsize}
\tablecaption{Possible identification of the features in the IRC spectra\tablenotemark{a}\label{table2}}
\tablehead{{Identification} & \colhead{Wavelength  ($\mu$m)} & \colhead{Remarks}}

\startdata

H\,I Br$\beta$ & 2.6259 & blended with H\,I Pf13\\
H$_2$ 1$-$0 O(2) & 2.6269 & marginally seen in H18.14.0\\
H\,I Pf13 & 2.6751 & blended with H\,I Br$\beta$ and not separately seen in the present spectra\\
H\,I Pf12 & 2.7583\\
H$_2$ 1$-$0 O(3) & 2.8031 & marginally seen in H18.14.0\\
H\,I Pf$\eta$ & 2.8730 \\
H$_2$O ice & 2.7$-$3.3 & absorption feature\\
H$_2$ 1$-$0 O(4) & 3.0039 & marginally seen in H18.14.0\\
H\,IPf$\epsilon$ & 3.0392 \\
PAH band & 3.29 & blended with H\,I Pf$\delta$\\
PAH band & 3.41 \\
PAH band & 3.48 & multicomponent \citep{geballe89} \\
H\,I Pf$\delta$ & 3.2970 & blended with the PAH 3.29\,$\mu$m and not  separately seen in the present spectra\\
H\,I Pf$\gamma$ & 3.7406 \\
H$_2$ 0$-$0 S(13) & 3.8468 \\
H$_2$ 0$-$0 S(12) & 3.9947 & blended with H\,I Br$\alpha$ and  not  separately seen in the present spectra\\
H\,I Hu14 & 4.0209 & blended with H\,I Br$\alpha$ and  not  separately seen in the present spectra\\
H\,I Br$\alpha$ & 4.0523 & blended with H$_2$ 0-0 S(12) and H\,I Hu14 \\
H\,I Hu13 & 4.1708 & blended with H$_2$ 0$-$0 S(11) \\
H$_2$ 0$-$0 S(11) & 4.1813 & blended with H\,I Hu13 \\
CO$_2$ ice & 4.27 & absorption feature\\
He\,I ($^3S_1$$-$$^3P_0$) & 4.2954 \\
H\,I Hu12 & 4.3765 & blended with H$_2$ 0$-$0 S(10) \\
H$_2$ 0$-$0 S(10) & 4.4099 & blended with H\,I Hu12 \\
H\,I Pf$\beta$ & 4.6538 & blended with H\,I Hu$\epsilon$ and H$_2$ 0$-$0 S(9) \\
H\,I Hu$\epsilon$ & 4.6725 & blended with H\,I Pf$\delta$  and H$_2$ 0$-$0 S(9) \\
H$_2$ 0$-$0 S(9) & 4.6946 &blended with Pf$\delta$  and H\,I Hu$\epsilon$ \\

\enddata
\tablenotetext{a}{
Not all the features are seen in the spectra (see text and remarks). }
\end{deluxetable}

\clearpage

\begin{deluxetable}{ccccccccc}
\tabletypesize{\scriptsize}
\rotate
\tablecaption{Intensities of the features\label{table3}}

\tablehead{& \multicolumn{2}{c}{Excess\tablenotemark{a}} &  \multicolumn{2}{c}{PAH band\tablenotemark{a}} & \colhead{H\,I Br$\alpha$\tablenotemark{a}}
& \multicolumn{3}{c}{Excess/PAH band\tablenotemark{b}}\\
\colhead{Object} & \colhead{4.4\,$\mu$m} & \colhead{4.65\,$\mu$m} &  \colhead{3.29\,$\mu$m} &  \colhead{3.42 + 3.48\,$\mu$m}
& & \colhead{aromatic} & \colhead{aliphatic} & \colhead{total}}

\startdata
Orion Bar & $13.6 \pm 1.1$ & $7.6\pm 1.3$ & $531 \pm 6$ & $192 \pm 3$ & $70.6 \pm 1.1$ 
& $0.026 \pm 0.002 $ & $0.040 \pm 0.007$ & $ 0.029 \pm 0.002$\\ 
M17          & $1.4 \pm 0.4$ & $3.2 \pm 0.6$  & $149 \pm 2$ & $48 \pm 1$  & $37.5\pm0.5$ 
& $0.009 \pm 0.003 $ &  $0.07 \pm 0.01 $ & $ 0.023 \pm 0.004$\\ 
G18.14.0 &$1.17 \pm 0.11$& $0.27 \pm 0.11$ & $47.8 \pm 0.6$ & $19.9 \pm 0.4$ & --- 
& $0.024 \pm 0.002 $ &  $0.014 \pm 0.006 $ & $ 0.021\pm 0.002$\\ 
\enddata
\tablenotetext{a}{in units of $10^{-18}$\,Wm$^{-2}$arcsec$^{-2}$. }
\tablenotetext{b}{The excess to PAH band intensity ratio.  `Aromatic' and `aliphatic' indicate the
ratio of the 4.4\,$\mu$m excess to the 3.29\,$\mu$m band intensities, and
that of the 4.65\,$\mu$m to the $3.42+3.48$\,$\mu$m band intensities, respectively, and `total'
shows the ratio of summation of the aromatic and aliphatic band intensities (see text).
}
\end{deluxetable}

\clearpage

\clearpage

\begin{thebibliography}{}
\bibitem[Allamandola et al.(1985)]{allamandola85}
 Allamandola, L. J., Tielens, A. G. G. M.. \& Barker, J. R.
1985, \apjl, 290, L25

\bibitem[Allamandola et al.(1989)]{allamandola89}
 Allamandola, L. J., Tielens, A. G. G. M.. \& Barker, J. R.
1989, \apjs, 71, 733

\bibitem[Bauschlicher et al.(1997)]{bauschlicher97}
Bauschlicher, C. W.,  Jr., Langhoff, S. R., Sandford, S. A., \& Hudgins, D. M.
1997, J. Phys. Chem. A., 101, 2414

\bibitem[Bica et al.(2003)]{bica03}
Bica, E., Dutra, C. M., Soares, J., \& Barbuy, B. 2003, \aap, 404, 223

\bibitem[Boesgaard \& Steigman(1983)]{boesgaard85}
Boesgaard, A. M., \& Steigman, G. 1983, \araa, 23, 319

\bibitem[Draine(2006)]{draine06} Draine, B. T. 2006, 
in ASP Conf. Ser. 348, Proc. Astrophysics in the Far Ultraviolet: Five 
Years of Discovery with FUSE, ed. G. Sonneborn, H. Moos, \& B-G Andersson
(San Francisco, CA: ASP) 58

\bibitem[Draine \& Li(2001)]{draine01}
Draine, B. T., \& Li, A. 2001, \apj, 551, 807

\bibitem[Ehrenfreund et al.(1996)]{ehrenfreund96}
Ehrenfreund, P., Boogert, A. C. A., Gerakines, P. A., et al. 1996, \aap, 315, L341

\bibitem[Epstein et al.(1976)]{epstein76}
Epstein, R. I., Lattimer, J. M., \& Shramm, D. N. 1976, \nat, 264, 198

\bibitem[Geballe et al.(1989)]{geballe89}
Geballe, T. R., Tielens, A. G. G. M., Allamandola, L. J., Moorhouse, A., \& Brand, P. W. J. L. 1989,
\apj, 341, 278

\bibitem[Gerakines et al.(1999)]{gerakines99}
Gerakines, P. A., Whittet, D. C. B., Ehrenfreund, P., et al. 1999, \apj, 522, 357


\bibitem[Gibb et al.(2004)]{gibb04}
Gibb, E. L., Whittet, D. C. B., Boogert, A. C. A., \& Tielens, A. G. G. M. 2004, \apjs, 151 35.

\bibitem[Hudgins et al.(1994)]{hudgins94}
Hudgins, D. M.,  Sandford, S. A., \&  Allamandola, L. J.
1994, J. Phys. Chem., 98, 4243

\bibitem[Hudgins et al.(2004)]{hudgins04}
Hudgins, D. M., Bauschlicher, C. W., Jr., \& Sandford, S. A. 
2004, \apj, 614, 770

\bibitem[Jura(1982)]{jura82}
Jura, M. 1982, in Advances in UV Astronomy: 4 Years of IUE Research, 
ed. Y. Kondo, J. M. Mead, \& R. D. Chapman (NASA CP 2238: Greenbelt MD:
NSAS), 54

\bibitem[Kaneda et al.(2009)]{kaneda09} Kaneda, H., Koo, B. C., Onaka, T., \& 
Takahashi, H.
2009, \pasj, 59, S401

\bibitem[Lee et al.(2011)]{lee11}
Lee, H.-G., Moon, D.-S., Koo, B.-C., et al. 2011, \apj, 740, 31

\bibitem[Leger \& Puget(1984)]{leger84}
Leger, A., \& Puget, J. L. 1984, \aap, 137, L5

\bibitem[Linsky et al.(2006)]{linsky06} Linsky, J. L. Draine, B. T., Moos, H. W.,
et al. 2006, \apj, 647, 1106

\bibitem[Mattila et al.(1996)]{mattila96}
Mattila, K., Lemke, D., Haikala, L. K., et al. 1996, \aap, 315, L353

\bibitem[Mazzitelli \& Moertti(1980)]{mazzitelli80}
Mazzitelli, I., \& Moretti, M. 1980, \apj, 235, 955

\bibitem[Messenger(2002)]{messenger02}
Messenger, S. 2011, \planss, 50, 1221

\bibitem[Mori et al.(2012)]{mori12} Mori, T. I., Sakon, I., Onaka, T., et al.
2012, \apj, 744, 68

\bibitem[Mori et al.(2013)]{mori13} Mori, T. I., Onaka, T., Sakon, I., Ishihara, D.,
Shimonishi, T., \& Bell, A. C. 2013, \apj, submitted

\bibitem[Murakami et al.(2007)]{murakami07} Murakami, H., Baba, H., Barthel, R., et al. 
2007, \pasj, 59, S369

\bibitem[Nummelin et al.(2001)]{nummelin01}
Nummelin, A., Whittet, D. C. B., Gibb, E. L., Gerakines, P. A., \& Chiar, J. E. 2001,
\apj, 558, 185

\bibitem[Ohyama et al.(2007)]{ohyama07} Ohyama, Y., Onaka, T., Matsuhara, H., et al.
2007, \pasj, 59, S411

\bibitem[Onaka et al.(1996)]{onaka96}
Onaka, T., Yamamura, I., Tanab\'e, T., Roellig, T. L., \& Yuen, L. 1996, \pasj, 48, L59

\bibitem[Onaka et al.(2007)]{onaka07} Onaka, T., Matsuhara, H., Wada, T., et al. 
2007, \pasj, 59, S401

\bibitem[Onaka et al.(2010))]{onaka10} 
Onaka, T., Matsuhara, H., Wada, T., et al. 2010, \procspie, 7731, 77310M


\bibitem[Peeters et al.(2004)]{peeters04} Peeters, E., Allamandola, L. J., Bauschlicher, C. W., Jr., 
et al. 2004, \apj, 604, 252

\bibitem[Pendleton \& Allamandola(2002)]{pendleton02}
Pendleton, Y. J., \& Allamandola, L. J. 2002, \apjs, 138, 75

\bibitem[Pettini \& Cooke(2012)]{pettini12} Pettini, M., \& Cooke, R. 2012, \mnras, 425, 2477

\bibitem[Pettini et al.(2008)]{pettini08} Pettini, M., Zych, B. J., Murphy, M. T., Lewis, A., 
\& Steidel, C. C. 2008, \mnras, 391, 1499

\bibitem[Prodanovi\'c et al.(2010)]{prodanovic10} Prodanovi\'c, T., Steigman, G., \& Fields, B. D. 
2010, \mnras, 406, 1108

\bibitem[Romano et al.(2006)]{romano06}
Romano, D., Tosi, M., Chiappini, C., \& Matteucci, F. 2006, \mnras, 369, 295

\bibitem[Shimonishi et al.(2010)]{shimonishi10}
Shimonishi, T., Onaka, T., Kato, D., et al. 2010, \aap, 514, A12

\bibitem[Storey \& Hummer(1995)]{storey95}
Storey, P. J., \& Hammer, D. G. 1995, \mnras, 272, 41

\bibitem[Tielens(1997)]{tielens97} Tielens, A. G. G. M. 1997, in AIP Conf. Proc. 402, 
Astrophysical Implications of the Laboratory Study of
Presolar Materials, ed. T. H. Bernatowicz and E. K. Zinner (New York: AIP), 523


\bibitem[Tielens(2008)]{tielens08} Tielens, A. G. G. M. 2008, \araa, 46, 289

\bibitem[Tosi(2010)]{tosi10} Tosi, M. 2010, in Proc. of IAU Symp. 268,
LIght Elements in the Universe, ed., C.Charbonnel, M. Tosi, F. Primas \& C. Chiappini,
153

\bibitem[Tsujimoto(2011)]{tsujimoto11} Tsujimoto, T. 2011, \mnras, 410, 2540

\bibitem[Tsumura et al.(2013)]{tsumura13}
Tsumura, K., Matsumoto, T., Matsuura, S., et al. 2013,
\pasj, in press (arXiv:1307.6736)

\bibitem[van den Ancker et al.(2000)]{vanden00}
van den Ancker, M. E., Tielens, A. G. G. M., \& Wesselium, P. R. 2000, \aap, 358. 1035

\bibitem[Verstraete et al.(1996)]{verstraete96}
Verstraete, L., Puget, J. L., Falgarone, E., et al. 1996,
\aap, 315, L337

\end{thebibliography}
\end{document}